\documentclass[conference]{IEEEtran}
\usepackage{lipsum}
\IEEEoverridecommandlockouts
\usepackage{cite}
\usepackage{amsmath,amssymb,amsfonts}
\usepackage{algorithmic}
\usepackage{graphicx}
\usepackage{textcomp}
\usepackage{xcolor}
\def\BibTeX{{\rm B\kern-.05em{\sc i\kern-.025em b}\kern-.08em
    T\kern-.1667em\lower.7ex\hbox{E}\kern-.125emX}}
\begin{document}

\title{Benchmarking Processor Performance by Multi-Threaded Machine Learning Algorithms\\
}

\author{\IEEEauthorblockN{\textsuperscript{} Muhammad Fahad Saleem}
\IEEEauthorblockA{\textit{Department of Computer Science} \\
\textit{National University of Computer and Emerging Sciences}\\
Islamabad, Pakistan \\
i192080@nu.edu.pk}
}

\maketitle

\begin{abstract}
Machine learning algorithms have enabled computers to predict things by learning from previous data. The data storage and processing power are increasing rapidly, thus increasing machine learning and Artificial intelligence applications. Much of the work is done to improve the accuracy of the models built in the past, with little research done to determine the computational costs of machine learning acquisitions. In this paper, I will proceed with this later research work and will make a performance comparison of multi-threaded machine learning clustering algorithms. I will be working on Linear Regression, Random Forest, and K-Nearest Neighbors to determine the performance characteristics of the algorithms as well as the computation costs to the obtained results. I will be benchmarking system hardware performance by running these multi-threaded algorithms to train and test the models on a dataset to note the differences in performance matrices of the algorithms. In the end, I will state the best performing algorithms concerning the performance efficiency of these algorithms on my system.
\end{abstract}


\section{Introduction}
\begin{center}
\begin{figure}[h]
\includegraphics[width=0.45\textwidth]{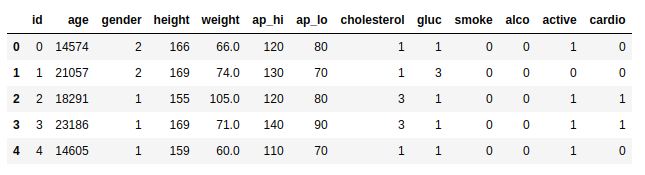}
\label{fig:1}
\caption{First five values of the dataset used}
\end{figure}
\end{center}

In the past decade, high-level research work in machine learning has transformed the technology, and now machine learning is emerging as one of the promising fields of interest. Machine Learning is being implemented in almost all fields and aspects of life, benefiting them in unimaginable ways. It outperforms previous practices bringing novelty, and thus promising the most significant insights into a brighter future. Data is an essential ingredient for machine learning algorithms as it impacts the efficiency of models. The data is primarily unstructured and must be refined by removing noisy data and cleansing it to extract meaningful information. While many researchers focus on increasing learning outcomes and accuracy improvements of machine learning algorithms, very little research is being done to decrease these calculations' computational cost. So, the use of Machine learning algorithms to evaluate, assess and benchmark the processor concerning performance should be the priority now for the researchers because the algorithms have come to the point that the accuracy is almost closer to the desired highest outcomes. 
The reasoning capability induced in computers by ML algorithms has led the computer to think and mimic humans is the most critical improvement in the technological world. Several programming languages are used to write ML algorithms, where Python is one of the influential languages due to the availability of rich libraries. This paper utilizes the Scikit-Learn library of Python to execute ML Clustering algorithms with the end goal of research.
My system has no OpenCL support, so I am also obliged to use only multi-threaded machine learning clustering algorithms and evaluate the performance of my system by running these algorithms on my system, engaging one to multiple threads. Multi-threading in machine learning helps reduce the execution time of programs where training and testing of datasets take much time. I will note the differences and improvements in the system's performance, such as the utilization of cores and execution time of Machine learning Algorithms.
I am using three Multi-Threaded Machine learning algorithms for clustering, which are listed below:
\begin{itemize}
\item Linear Regression
\item Random Forest
\item K-Nearest Neighbors
\end{itemize}
In my research work, I have used three different multi-threaded clustering algorithms whose results, in the end, are compared with each other by executing the algorithms on Cardio dataset that can be seen in Fig \ref{fig:1}. This research work aims to assess the performance of system hardware when these multi-threaded algorithms are running on it.
\subsection{Paper Organization}
 The rest of the Research Paper is organized as follows: Related work will be discussed in Section 2, Experimental Results will be displayed in Section 3, and Section 4 will present and discuss the results obtained from the experimentation. The acknowledgment will be in section 5, and the conclusion of my research work will be in the last section.

\section{Related Work}
I reviewed the previous research work on this topic and went through the following research papers related to computer processors' performance when running machine learning algorithms. I found out that related work has been done in the domain of architectural performance by the use of machine learning algorithms. In the paper \cite{penney2019survey}, the authors have applied machine learning algorithms to explore the system architecture by enabling diverse training methods and learning models to identify the effective adjustments between performance and overhead based on task requirements, which led to the categorization of system behavior based on multiple level details. In this paper \cite{sayadi2017machine}, the authors focused on the view that architectures need to optimize and fine-tune metric parameters to attain efficiency concurrently. They also pressed that the schedule challenges and suitable configurations are mandatory to achieve efficiency when running a multi-threaded algorithm. The optimal processor configuration at run time is dealt with by assessing the characteristics and optimality objective of the multi-threaded algorithm. Heterogeneous architectures are suitable for running Machine learning algorithms as they have the tuning optimization capability to improve the performance and energy consumption of the system. The model fetches hardware performance counters at run time from multi-threaded applications and predicts the most efficient performance configuration to set the number of threads and task frequency. The non-linear regression and neural network models provide more efficient system configuration than simple regression or decision tree-based models, increasing system design complexity. In this paper \cite{garcia2019estimation}, the authors have adopted energy as a metric in the case of machine learning algorithms. Previously, the research was focused on attaining high levels of accuracy, ignoring computational constraints. The authors have reviewed different approaches to estimate energy consumption when running machine learning algorithms. The absence of suitable tools to quantify and create models in existing ML suites has led to the absence of such metrics today. The authors have explored the fundamental approaches for energy consumption, mapping to ML algorithms. They described advanced energy consumption for data mining and convolutional neural networks. In this paper \cite{yu2017evaluating}, the authors have represented benchmarking suites for serial and parallel workloads to characterize architecture designs like Non-uniform memory access (NUMA), Simultaneous multi-threading (SMT), and Turbo boost concerning power and energy consumption and efficiency. In this paper \cite{nemirovsky2017machine}, authors have stressed that due to the abundance of the heterogeneous system, there is a need to develop new CPU scheduling techniques having the capability of exploring the diversity of computational resources. Latest machine learning algorithms, including artificial neural networks, are helpful in prediction models for various fields. So this can be helpful to pioneer architectural designs capable of enhancing mapping and system throughput. So, it is required to work on novel models supporting huge hardware diversity.

I have selected three machine learning classification algorithms: Linear Regression, Random Forest, and K-Nearest Neighbors. I will run these algorithms employing multiple threads ranging from one to four and assess the differences in execution time and processing speed when running the code. I have selected a prominent dataset \cite{ulianova_2019} related to health from Kaggle. It has 13 columns and  70,000 entries of cleaned data split as 75 percent for training and 25 percent for testing purposes. I will plot the results of the algorithms using Matplot libraries. In the end, I will be benchmarking my system concerning the performance of my processor while employing multiple threads when running the Machine learning algorithms. These three Multi-threaded machine learning classification algorithms will be evaluated by running the code on single-core and multiple cores of the processor to compare their performance with each other.

\section{Experimental Details}
This section consists of three parts where first part has the code of algorithms used, the second part consists of the testbed, and the third part will contain the results. The code part will have the main code of the clustering algorithms. The testbed will state the tools and technologies used along with the environment required. The results part will show the expected results obtained from the experimentation.

\subsection{Test Bed}
We have the following items in our test bed
\begin{itemize}
    \item Multi-Threaded Clustering algorithms
    \item Cardio Dataset
    \item Linux OS
    \item Python3
    \item Perf Tool
    \item Valgrind Tool
    
\end{itemize}

\subsection{Code and properties of Algorithms}
	
\subsubsection{Linear Regression}
The following is the code for the Linear Regression algorithm, which employs four threads. 
\\
\begin{figure}[htbp]
\centerline{\includegraphics[width=.4\textwidth]{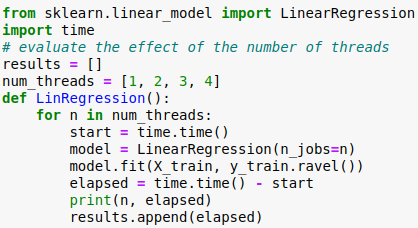}}
\caption{Linear Regression code }
\label{fig:2}
\end{figure}
\newpage
line profiler assesses the properties of code line by line as you can see that it has pointed that majority if the computation time consumed in Random forest code is during the model fitting part of the code which is 99.8\%.
\begin{figure}[htbp]
\centerline{\includegraphics[width=.4\textwidth]{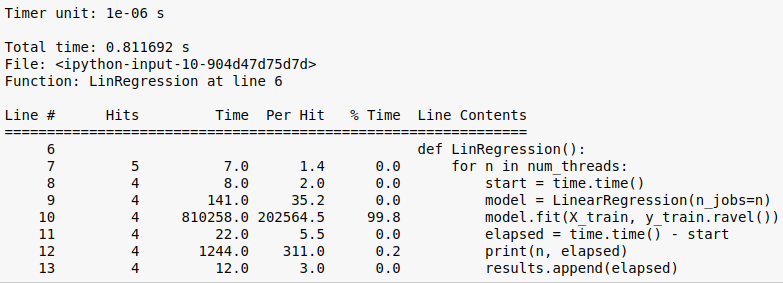}}
\caption{LR Line Profiler }
\label{fig:3}
\end{figure}
\\
LR execution time reduces with the increase in No. of threads.
\begin{figure}[htbp]
\centerline{\includegraphics[width=.3\textwidth]{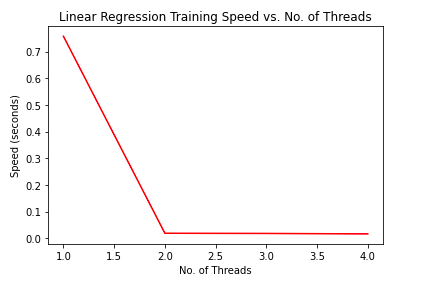}}
\caption{LR plot }
\label{fig:4}
\end{figure}
\\

\subsubsection{Random Forest }
The following is the code for Random Forest algorithm that employs four number of threads. 
\begin{figure}[htbp]
\centerline{\includegraphics[width=.4\textwidth]{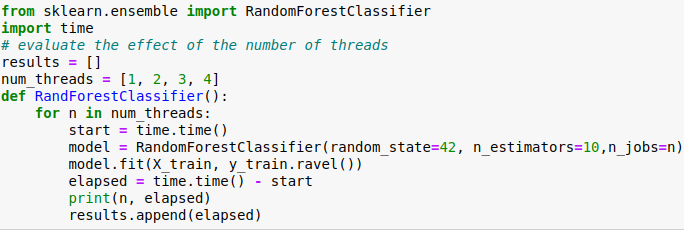}}
\caption{Random Forest code }
\label{fig:5}
\end{figure}
\\
The line profiler assesses the properties of code line by line as you can see that it has pointed that the majority of the computation time consumed in Random forest code is during the model fitting part of the code, which is 99.9\%.
\begin{figure}[htbp]
\centerline{\includegraphics[width=.4\textwidth]{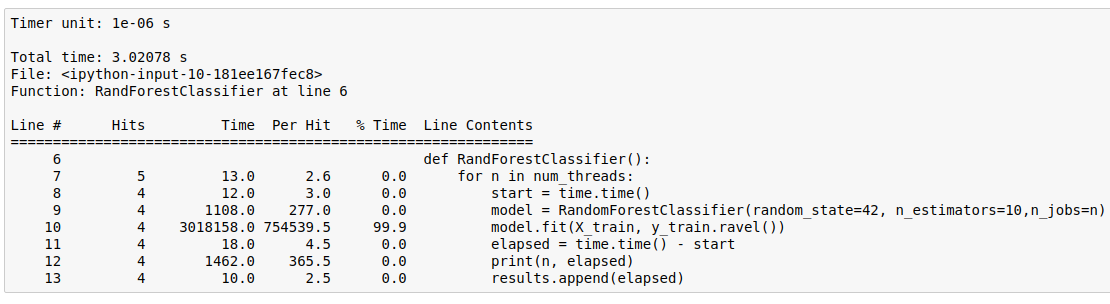}}
\caption{RF Line Profiler }
\label{fig:6}
\end{figure}
\\
\newpage
RF execution time also reduces with the increase in No. of threads.
\begin{figure}[htbp]
\centerline{\includegraphics[width=.3\textwidth]{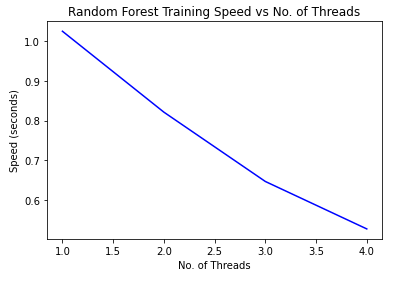}}
\caption{RF plot }
\label{fig:7}
\end{figure}

\subsubsection{K-Nearest Neighbors}
The following is the code for K-Nearest Neighbors algorithm that employs four number of threads.
\begin{figure}[htbp]
\centerline{\includegraphics[width=.4\textwidth]{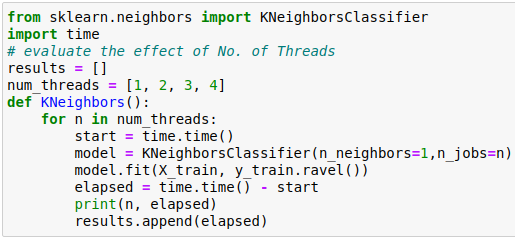}}
\caption{K-Nearest Neighbors code }
\label{fig:8}
\end{figure}
\\
99.7\% of time is consumed during the model fitting part of the KNN code
\begin{figure}[htbp]
\centerline{\includegraphics[width=.4\textwidth]{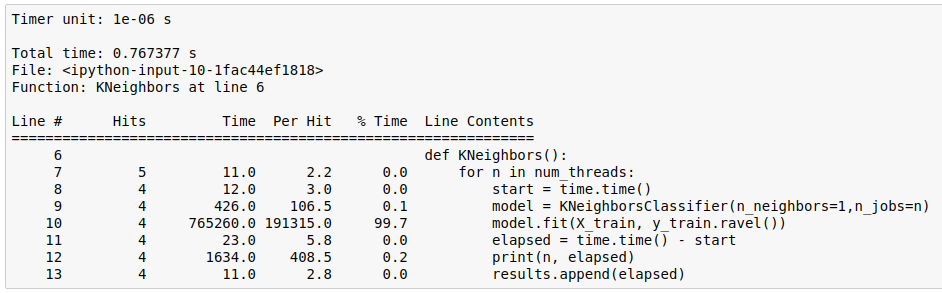}}
\caption{KNN Line Profiler }
\label{fig:9}
\end{figure}
\\
KNN execution time show variant properties with the increase in No. of threads.
\begin{figure}[htbp]
\centerline{\includegraphics[width=.3\textwidth]{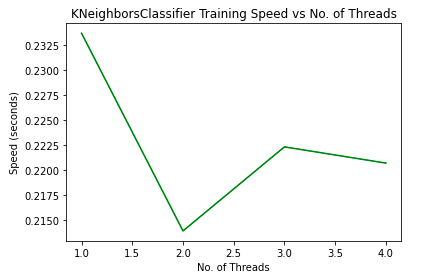}}
\caption{KNN plot }
\label{fig:10}
\end{figure}
\newpage
\subsection{Results}
I have executed the above three classification algorithms, namely Linear Regression, Random Forest, and K-Nearest Neighbors, on four threads using the Cardio dataset. Then I have shown the results in section 4 of the paper.

\subsection{Hardware and Software Specifications} 
The following are the hardware specifications of the system which we will be using in this research work: 
\begin{table}[htbp]
\caption {Hardware Specifications Of Machine} \label{tab:title1} 
\begin{tabular}{|c|c|}
\hline
Processor Name :&  Intel(R) Core i5 (R) 2430M @ 2.40GHz \\\hline
Model Number: & 2430M \\\hline
Memory Configuration: & Described Below\\\hline
Size:&  6 GB \\\hline
Type: & DDR3/L/-RS 1333/1600\\\hline
Speed: & 2.40 GHz\\\hline
Core/Thread:&  2/4\\\hline
Cache Configuration:&  Describe Below\\\hline
Cache: & L1 data L1 instruction L2 L3\\\hline
Size: & 2 x 32 KB 2 x 32 KB 2 x 256 KB 3 MB\\\hline
Associativity: &  8-way set associative \\\hline
Line size: & 64 bytes \\\hline
Hard Disk:     & Solid State Drive 250 GB \\\hline
\end{tabular}
\end{table}
\subsection{Software Specifications} 
The following are the software specifications of the System that we are using in this research work: 

\begin{table}[htbp]
\caption {Software Specifications Of System} \label{tab:title} 
\begin{tabular}{|c|c|}
\hline
Operating System:    & Ubuntu 16.04.7 LTS (Xenial Xerus) \\ \hline
Kernel Version: & 4.15-generic \\ \hline
OS Architecture:     & 64 Bit\\ \hline
Performance Monitoring Tool:           & Perf Version 5.10.1                                    \\ \hline
Programming Language    &  Python   
 \\ \hline
\end{tabular}
\end{table}

\section{Results And Discussion}
I have used \textbf{Perf }command, which is a performance measurement and analysis tool that runs in Linux operating system. The Perf command runs on the Linux terminal. I have also used \textbf{Valgrind} to check for memory leaks if there are any to assess the efficiency of the algorithms. The commands that I have used are as follows: 
\\
\begin{itemize}
\item \textbf{perf stat ./filename.py}
\item \textbf{perf stat -e cpu-clock, branch-instructions ./aca.py} 
\item \textbf{perf,branch-loads, branch-load-misses ./aca.py}
\item \textbf{perf cache-references, cache-misses ./aca.py}
\item \textbf{perf L1-dcache-loads, L1-dcache-load-misses ./aca.py}
\item \textbf{perf L1-dcache-stores L1-icache-load-misses ./aca.py}
\item \textbf{valgrind --leak-check=yes ./aca.py}

\newpage
\end{itemize}
\subsection{Technical Terms Used}
This section will describe some of the technical terms used in the research paper to make it easier to understand the results.
\\
\subsubsection{Branches} 
A branch is a predicted set of routine walks that your computer's processor might want to take, so you take care of it by making preparations in advance to save time and increase the system's latency.
\\
\subsubsection{Branch Misses}
The concept of an instruction-level pipeline architecture is that whenever you predict wrong for system behavior when it has to take a branch, and it doesn't, this is called a branch miss.
\\
\subsubsection{Stalls}
Stalls are created whenever a hazard has occurred, which means that at a given instance of time, two or more clients are accessing the same data, which will create a deadlock. 
\\
\subsubsection{Context Switching}
It is switching programs from running the state into waiting and waiting for the state to running to give an impression that the system is achieving parallelism. Parallelism in architecture is the main demand in systems nowadays. However, today, parallelism is achieved by giving the impression that the processes are running simultaneously, which is achieved by frequent context switching. \\
\subsubsection{CPU Migration}
it is the shuffling of the programs which are running and are being processed onto the multiple cores. The scheduler manages it without the program knowing it.
\\
\subsubsection{Page Faults}
Modern operating systems have virtual memories where processes are loaded in the first place when they are going to be executed. Virtual memories have fix size partitions called pages. Whenever the process loads onto virtual memory, a page copy is also loaded into physical memory. In Physical Memory, there are fixed-size partitions called frames. If the required page of a process is not available in the frame at a specific time, then a page fault occurs, thus generating an error.
\\
\subsubsection{Time Consumed}
it is the time that a program or algorithm takes to complete its execution
\\
\subsection{Perf Results}
 I have applied the Perf command on the .py version of my Code files that employed Cardio data set having 70,000 values to assess the efficiency of my system by running the Multi-Threaded Machine learning Clustering Algorithms. 
\\
\newpage
\subsubsection{\textbf{Execution time Of Clustering algorithms}}
When I executed the three clustering algorithms employing the different number of threads ranging from one to four using the cardio dataset, I got the execution time of Linear Regression, Random Forest, and K-Nearest Neighbors 0.03, 0.549, and 0.165 seconds. So, these results show that Linear Regression worked efficiently in terms of time when considering the program's execution time.

\begin{figure}[htbp]
\centerline{\includegraphics[width=.4\textwidth]{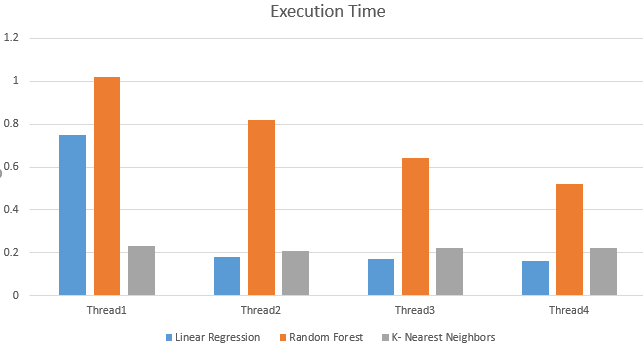}}
\caption{Execution time of Multi-threaded Clustering Algorithms }
\label{fig:11}
\end{figure}

\subsubsection{\textbf{Cycle Results}}
When i executed the three clustering algorithms on cardio dataset i got the results of No. of Cycles in Linear Regression, Random Forest and K-Nearest Neighbors as 7327945876, 18029911435 and 6523415458. So, these results show that KNN worked fine when considering the number of Cycles in the program.
\begin{figure}[htbp]
\centerline{\includegraphics[width=.3\textwidth]{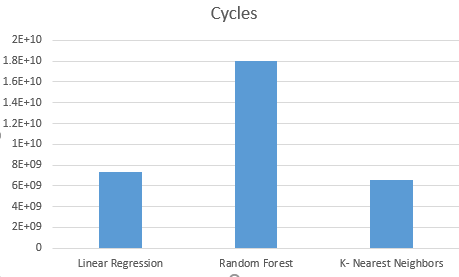}}
\caption{Cycle results }
\label{fig:12}
\end{figure}
\subsubsection{\textbf{Instruction Results}}
When i executed the three clustering algorithms on cardio dataset i got the results of  of Linear Regression, Random Forest and K-Nearest Neighbors as 4496839433, 17305738289 and 5472318307. So, these results show that Linear Regression worked fine when considering the No. of Instructions in the program.
\begin{figure}[htbp]
\centerline{\includegraphics[width=.3\textwidth]{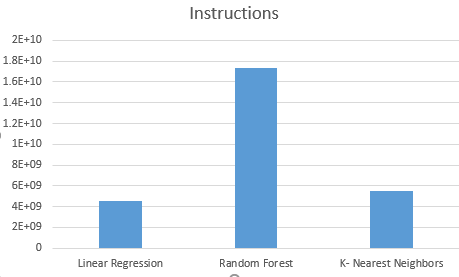}}
\caption{Instruction results }
\label{fig:13}
\end{figure}
\newpage
\subsubsection{\textbf{Branch Results}}
When I executed the three clustering algorithms on the cardio dataset, I got the results of No. of branches for Linear Regression, Random Forest, and K-Nearest Neighbors as 947590165, 3034816910, and 1001993414. So, these results show that Linear Regression worked fine when considering the No. of branches in the program.
\begin{figure}[htbp]
\centerline{\includegraphics[width=.3\textwidth]{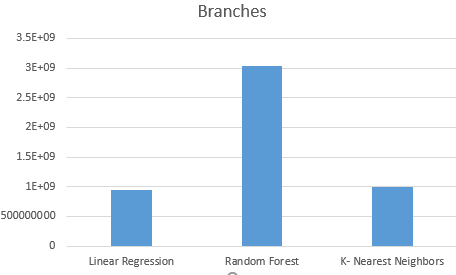}}
\caption{Branch results }
\label{fig:14}
\end{figure}
\\
\subsubsection{\textbf{Branch Misses Results}}
When I executed the three clustering algorithms on the cardio dataset, I got the results of Branch Misses in Linear Regression, Random Forest, and K-Nearest Neighbors as 29615147, 155008845, and 23223335. So, these results show that KNN worked fine when considering the Branch Misses in the program.
\begin{figure}[htbp]
\centerline{\includegraphics[width=.3\textwidth]{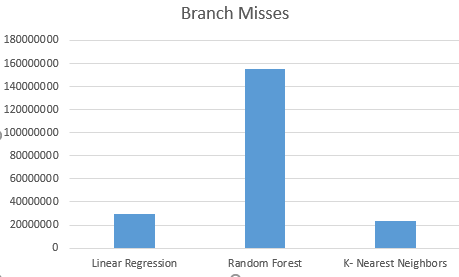}}
\caption{Branch Misses result }
\label{fig:15}
\end{figure}

\subsubsection{\textbf{Stall Results}}
When i executed the three clustering algorithms on cardio dataset i got the results of Stalls in Linear Regression, Random Forest and K-Nearest Neighbors as 5062744819, 10506871966 and 4186123236. So, these results show that KNN worked fine when considering the Stalls in the program.
\begin{figure}[htbp]
\centerline{\includegraphics[width=.3\textwidth]{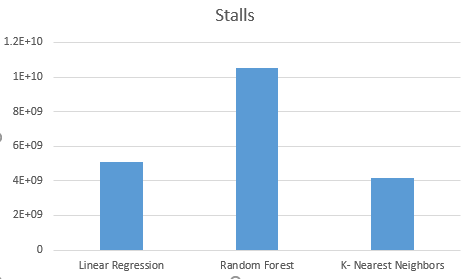}}
\caption{Stalls Results}
\label{fig:16}
\end{figure}
\newpage
\subsubsection{\textbf{Context Switches Results}}
When i executed the three clustering algorithms on cardio dataset i got the results of Context Switches in Linear Regression, Random Forest and K-Nearest Neighbors as 2430, 2286 and 158. So, these results show that KNN worked fine when considering the Context Switches in the program.
\begin{figure}[htbp]
\centerline{\includegraphics[width=.3\textwidth]{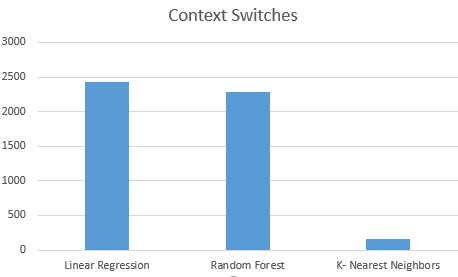}}
\caption{Context Switching results}
\label{fig:17}
\end{figure}
\subsubsection{\textbf{CPU Migration Results}}
When i executed the three clustering algorithms on cardio dataset i got the results of  of Linear Regression, Random Forest and K-Nearest Neighbors as 26, 79 and 1. So, these results show that KNN worked fine when considering the CPU Migrations in the program.
\begin{figure}[htbp]
\centerline{\includegraphics[width=.2\textwidth,height=6cm]{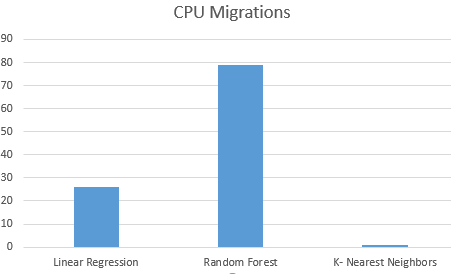}}
\caption{CPU Migration results}
\label{fig:18}
\end{figure}
\subsubsection{\textbf{Page Faults Results}}
When i executed the three clustering algorithms on cardio dataset i got the results of Page Faults in Linear Regression, Random Forest and K-Nearest Neighbors as 46397, 62833 and 44676. So, these results show that KNN worked fine when considering the Page Faults in the program.
\begin{figure}[htbp]
\centerline{\includegraphics[width=.2\textwidth]{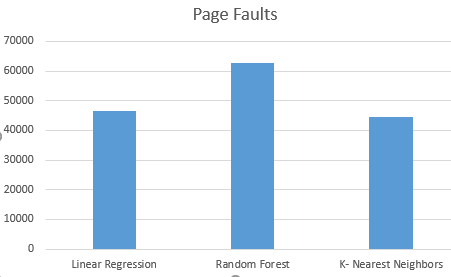}}
\caption{Page Faults results}
\label{fig:19}
\end{figure}
\newpage
\subsubsection{\textbf{Time Consumed Results}}
When I executed the three clustering algorithms on the cardio dataset, I got the time to display Linear Regression, Random Forest, and K-Nearest Neighbors as 11.768, 4.56, and 1.627. So, these results show that KNN worked fine when considering the time consumption of the program.
\begin{figure}[htbp]
\centerline{\includegraphics[width=.3\textwidth]{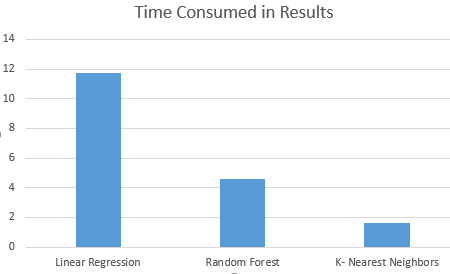}}
\caption{Time Consumed results}
\label{fig:20}
\end{figure}
\\
So, by comparing and assessing the differences in results of the three algorithms, I have concluded that K- Nearest Neighbors work efficiently when the machine's performance is kept in mind. However, the execution time of Linear regression is less when compared to the other two clustering algorithms. In addition to that, Random forest is consistent in reducing time with the increase in the number of threads as the other two algorithms have outliers in performance concerning execution time when the no. of threads keeps on increasing.
\\
\section{Acknowledgment}
I would like to thanks Allah Almighty for giving me strength, enough power, and knowledge to complete this research work. I have learned a lot in the whole process of research, which will for sure help me in my future. I want to thank Dr. Arshad Islam, department of CS, FAST NUCES Islamabad, who guided me in learning and doing quality research work. The learned skills have helped me in complex situations and difficult times during the research. In the end, I would like to thank all my friends and family members for their help, support, and encouragement.
\\
\section{Conclusion}
Machine learning is becoming an essential part of our technological world. While much of the work is being done to improve ML models' accuracy, research should also be done on the computation costs to these models and should try to minimize the costs. I have performed the performance analysis of my system when running Machine learning algorithms and have analyzed the performance of my system when running ML algorithms. In future work, Machine learning algorithms should be utilized to enhance the performance capability of different systems by exploring the architecture and then fine-tuning them to improve system competence.
 \newpage
\bibliography{Reference}{}
\bibliographystyle{plain}
\end{document}